# Fighting Against XSS Attacks:
## A Usability Evaluation of OWASP ESAPI Output Encoding

Chamila Wijayarathna
University of New South Wales
Australia
c.diwelwattagamage@student.unsw.edu.au

Nalin A. G. Arachchilage
University of New South Wales
Australia
nalin.asanka@adfa.edu.au

## Abstract

*Cross Site Scripting (XSS) is one of the most critical vulnerabilities exist in web applications. XSS can be prevented by encoding untrusted data that are loaded into browser content of web applications. Security Application Programming Interfaces (APIs) such as OWASP ESAPI provide output encoding functionalities for programmers to use to protect their applications from XSS attacks. However, XSS still being ranked as one of the most critical vulnerabilities in web applications suggests that programmers are not effectively using those APIs to encode untrusted data. Therefore, we conducted an experimental study with 10 programmers where they attempted to fix XSS vulnerabilities of a web application using the output encoding functionality of OWASP ESAPI. Results revealed 3 types of mistakes that programmers made which resulted in them failing to fix the application by removing XSS vulnerabilities. We also identified 16 usability issues of OWASP ESAPI. We identified that some of these usability issues as the reason for mistakes that programmers made. Based on these results, we provided suggestions on how the usability of output encoding APIs should be improved to give a better experience to programmers.*

## 1. Introduction

2017 OWASP Top 10 report [1] listed Cross Site Scripting (XSS) as one of the 10 most critical security risks for web applications[1]. XSS has been ranked among the top 10 most critical security risks for web applications since the start of OWASP Top 10 project in 2010. According to OWASP Top 10 2017 report [1], XSS is the 2nd most prevalent security issue and it exists in 2/3 of web applications.

XSS is a security vulnerability that allows attackers to inject client-side malicious scripts into web pages of applications. Those scripts will thereafter execute in victims' web browsers when they access those web pages [2]. Successful XSS attacks can result in serious security violations for both the web site and the user. An attacker can inject a malicious code into user input of a web application, and if the input is not validated, the code can steal cookies and login credentials [3, 4, 5], transfer private information [4, 5], hijack a user's account [3, 4, 5], manipulate the web content [4, 5], cause denial of service [6] and many other malicious activities [2, 7].

A recent XSS vulnerability identified in eBay allowed attackers to include malicious JavaScripts into auction description field of a selling item [3]. When a genuine user visits a listing with a such malicious auction description, the attached JavaScript will automatically execute. Some attackers have reportedly include scripts that transfer login details of users, so when a user visits the affected listing, the script will be executed and it will transfer user's login details/cookies to the attacker [3].

Programmers can prevent their web applications from being vulnerable to XSS attacks by separating untrusted data from active browser content [1]. One of the techniques for achieving this is encoding untrusted output data when loading them into browser content [1, 8]. However, most programmers who are involved in application development are not security experts and they are not capable of implementing such security techniques on their own [9]. Therefore, security experts have developed these functionalities so that non-expert programmers can use these functionalities when developing applications via Application Programming Interfaces (APIs). There are several APIs that implement output encoding functionalities that programmers can use to protect their applications from being vulnerable to XSS attacks. OWASP Enterprise Security API (ESAPI)[2], OWASP Java

[1]OWASP - Open Web Application Security Project (https://owasp.org/)

[2]https://www.owasp.org/index.php/Category:OWASP_Enterprise_Security_API

Encorder[3] and Microsoft Anti-Cross Site Scripting Library[4] are a few such security APIs.

However, 2/3 of applications still being vulnerable to XSS attacks [1] implies that programmers have failed to effectively use functionalities exposed by these APIs to prevent their applications from being vulnerable to XSS attacks. One possible reason for this can be the lack of usability of APIs that provide output encoding functionalities. When APIs that provide output encoding are not usable, programmers will fail to correctly use them in their code and therefore, will fail to protect applications they develop from being vulnerable to XSS attacks. Previous research has shown that less usable APIs, especially those that provide security functionalities, result in programmers incorrectly using them and introduces security vulnerabilities to applications they develop [10, 11, 12].

In this study, we tried to evaluate the usability of one of the most commonly used APIs that provide output encoding functionalities, OWASP ESAPI, and attempted to identify how usability issues of OWASP ESAPI would fail programmers who want to fix XSS vulnerabilities in their applications. To achieve this objective, we conducted a qualitative experimental study with 10 programmers. In the experiment, programmers used OWASP ESAPI to fix XSS vulnerabilities in a web application. The study employed think-aloud method [13] and cognitive dimensions questionnaire method [14] to identify usability issues they encounter while performing this task.

From the data we gathered, we identified 16 usability issues of OWASP ESAPI. We also found 3 types of mistakes that programmers made while completing the task that caused them to fail in successfully fixing the XSS vulnerability. We identified that some of the identified usability issues are a main reason for mistakes that programmers made. Based on this, we provided suggestions to improve the usability of security APIs that provide output encoding functionalities.

The paper is organized as follows. Section 2 reviews previous related research. Section 3 describes the experiment methodology and section 4 presents the findings of the study. In section 5, we discuss the findings and provide suggestions to improve usability of output encoding APIs. Finally, we conclude the paper with discussion of limitations and conclusion.

[3] https://www.owasp.org/index.php/OWASP_Java_Encoder_Project

[4] https://www.microsoft.com/en-au/download/details.aspx?id=43126

## 2. Related Work

The relationship between the usability of security APIs (APIs that provide security related functionalities) and security of end user applications that use those security APIs has become a topic of high interest among researchers recently [14, 15, 16, 17, 18, 19]. There have been several studies that discuss and investigate this relationship [9, 15, 16, 19, 20].

By pointing that programmers are not security experts, Wurster and van Oorschot argue that improving usability of tools and APIs that programers use is important to minimize mistakes they make while developing applications [9]. Acar et al. [20] also highlight the importance of the usability of security APIs by pointing that programmers who make use of security APIs are not experts of security. Mindermann [19] argues that security of an application will be far better if the libraries and APIs used to develop that application are more usable. He stresses the importance of applying usability research for security APIs to deliver more usable security APIs.

Even though the importance of the usability of security APIs has been identified and discussed [9, 15, 16, 19], not much work has been done to evaluate the usability of security APIs. Gorski and Iacono [17] presented 11 characteristics that need to be considered when evaluating the usability of security APIs. Green and Smith [18] introduced 10 rules for developing usable security APIs. By considering these 2 sets of guidelines and by referring to previous work done on usability evaluation of general APIs, Wijayarathna, Arachchilage and Slay [14] presented a cognitive dimensions framework, which consists of 15 dimensions to be used in the usability evaluation of security APIs.

There are few studies that involve empirical evaluations of security APIs [15, 21]. Acar et al. [15] evaluated and compared usability of 5 cryptographic APIs for python. They identified that security of applications that use security APIs is significantly related to the usability of security APIs used for developing the application. Wijayarathna and Arachchilage [21] used think-aloud approach and cognitive dimensions questionnaire method to identify usability issues of Bouncycastle API, an API that provides cryptographic functionalities.

There is a huge body of research done on the field of XSS attacks [7, 22, 23, 24, 25, 26, 27, 28, 29]. Previous research has proposed and discussed various methods for XSS attack implementation [22, 23], XSS attack detection [24, 25], XSS attack prevention [26, 27] and XSS vulnerability detection [28, 29]. Various methods have been proposed to prevent XSS attacks by developing applications that are not vulnerable to XSS attacks [7, 26, 27]. However, the most commonly practiced method for XSS prevention is the output encoding of untrusted data [8]. Even though so many areas related to XSS have been investigated, as per authors knowledge, there have been no investigation on the usability of output encoding APIs and how usability issues of output encoding APIs result in applications being vulnerable to XSS attacks. Our work attempts to fill this gap by studying programmers who use OWASP ESAPI to secure web applications from XSS vulnerabilities.

## 3. Methodology

The study was designed to identify usability issues of OWASP ESAPI that programmers encounter while they are using it to protect their applications from XSS vulnerabilities. Furthermore, we intended to observe how usability issues affect programmers and security of applications they develop. This study was approved by the Human Research Ethic Committee of our university.

Conducting a user study is a widely known method for identifying usability issues of APIs [14, 15, 21]. In a user study based usability evaluation, evaluators will recruit programmers and ask them to complete some tasks that will require them to use the API under evaluation. Then, evaluators will identify usability issues by observing programmers while they are completing the task and from the feedback they give upon the completion of the task.

We employed two techniques to identify usability issues that programmers encounter while using the API, which are popular in the API usability community. Those are:

- Cognitive dimensions questionnaire based method [14, 21, 30]
- Think-aloud method [13, 21]

Cognitive dimensions questionnaire based method [14, 30, 21] is the only methodology that has been proposed to specifically use in evaluating the usability of security APIs [14]. The cognitive dimensions framework presents a set of dimensions that describe aspects of a tool or an API that impact its usability [14, 31]. We used the version of cognitive dimensions framework proposed by Wijayarathna, Arachchilage and Slay [14] in this study, which consists of 15 cognitive dimensions. This framework is embedded to the usability evaluation process through the cognitive dimensions questionnaire [14, 31]. In the evaluation process, once participant programmers complete a programming task, they have to individually answer the questionnaire based on their experience [14, 31]. In this way, evaluators can evaluate each aspect of the API that is covered by the cognitive dimensions framework and identify usability issues of the API.

We used think-aloud method [13] to get more insights into the issues that were identified by the cognitive dimensions questionnaire method and how those issues affected programmers. Using two different techniques helps to improve the reliability of data we collect [32]. Furthermore, we expected that it would help us to identify a broad range of usability issues of the API.

### 3.1. Task Design

First we had to design a programming task for participants to follow.

We designed a programming task where participant programmers have to use output encoding functionalities provided by OWASP ESAPI to fix XSS vulnerabilities in a Java servelet web application. We developed a simple online forum type web application that allows users (end-users of the application) to create new forum posts by entering text data as input, stores that data in a back-end data store and shows the forum posts in the web interface when requested by end-users. Web pages that show forum posts loaded data that were entered by end-users (hence untrusted) into web page content without performing any encoding, hence making the application vulnerable to stored XSS attacks [2]. Appendix A shows the source of a sample page in the application where untrusted data entered by users loaded into active HTML, HTML attribute and JavaScript contents. The provided application contained two web pages with XSS vulnerabilities in different type of elements (i.e. HTML, HTML attribute and JavaScript). We asked participants to locate places in the code with XSS vulnerabilities and fix them using output encoding functionalities provided by OWASP ESAPI.

### 3.2. Pilot Study

Before conducting the main study, we requested 3 participants, who are known to the first author and not related to the study, to complete the task by following the guidelines to verify whether guidelines are clear and convey the expected meaning. They also answered the cognitive dimensions questionnaire proposed by Wijayarathna, Arachchilage and Slay [14] after completing the task. We did minor modifications to task guidelines and the questionnaire (change wording so participants can better understand task guidelines and questions) based on their results. Modified versions were used in the main study.

### 3.3. Participants

We recruited programmers with Java experience from GitHub to participate in the study. We used Github to recruit participants rather than recruiting participants from our university or local software development firms, to get a more diverse sample of programmers. This is a widely accepted and used method among researchers to recruit participants for developer studies [15, 16]. Furthermore, recruiting participants from Github helps to get participants with more experience in software development, which improves the ecological validity of the study [16]. We extracted publicly available email addresses of Java developers with significant contributions to Java projects and sent emails inviting them to participate in our study. We offered them with a $15 Amazon gift voucher as a token of appreciation for the participation. In the invitation email, we included a link to sign up for the study. Furthermore, we informed them that participation is voluntary and participants can withdraw from the study at any time. Sign up form required participants to enter their name and email address, which were required to send study material to them. However, such personally identifiable information of the participants were removed from the final data set which we used for the analysis.

We conducted 4 usability studies parallelly and recruited participants to all 4 studies together. We sent 13000 invitations to Java developers and 347 developers signed up for the study by completing the sign up form. Some emails we sent were bounced and some developers requested to be removed from our list, a request we honored. Furthermore, some people replied back to us saying that they are unable to participate in the study. Once people signed up, we filtered out those who did not have any software development experience since our target sample for the study was software developers. Furthermore, we filtered out participants with no experience in using Java because if a participant faces issues with programming language while completing the task, we may not be able to clearly identify usability issues of the API they had come up with. Then we divided participants who signed up into four studies we conducted based on their demographics. We selected 51 programmers for the ESAPI API study and sent study material for them. However, some of them informed that they are not able to complete the study and some of them dropped out without informing us. A total of 10 participants completed the study. Table 1 summarizes demographics of the participants that took part in the study.

**Table 1. Participant demographics summary**

| Demographic | Number | Percent |
|---|---|---|
| **Software Development Experience** | | |
| Less than 1 year | 2 | 20% |
| 1 to 3 years | 6 | 60% |
| 5 to 10 years | 1 | 10% |
| more than 10 years | 1 | 10% |
| **Java Experience** | | |
| 1 to 2 years | 1 | 10% |
| 2 to 3 years | 4 | 40% |
| 3 to 5 years | 2 | 20% |
| more than 5 years | 3 | 30% |
| **Number of Hours Spending for Java programming** | | |
| Currently Not using Java | 2 | 20% |
| 1 to 10 hours per week | 3 | 30% |
| 11 to 20 hours per week | 3 | 30% |
| 21 to 30 hours per week | 2 | 20% |
| **Participant has previously used OWASP ESAPI or not** | | |
| Yes | 2 | 20% |
| No | 8 | 80% |

### 3.4. Study Procedure

Participants completed the task remotely on their own computers and we suggested them to complete the task in a time comfortable to them. We requested them to think-aloud [13] and record their screens with voice (so thinkaloud results will be recorded) while completing the task. Once participants completed the task, they were asked to send their source codes with video recordings to us via email. Participants spent 35 minutes in average to complete the task. Then each participant had to complete the cognitive dimensions questionnaire [14, 21, 30], which we shared with them via Google forms.

### 3.5. Identification of Usability Issues

Once we finished the data collection, source codes of solutions that participants developed were evaluated to see whether they have fixed the XSS vulnerability correctly. Application had 3 types of contexts (HTML, HTML Attribute and JavaScript) where participants had to protect using 3 different methods of the API (*encodeForHTML(), encodeForHTMLAttribute(), encodeForJavaScript()*). We separately evaluated whether participants had protected these elements of the web application successfully.

Then analysis of video recordings and questionnaire responses were done manually by one analyst. We used manual analysis since our data set was small [33]. Questionnaire answers were analysed prior to analysing videos and identified usability issues that exist in OWASP ESAPI. After analysing questionnaire answers, recordings were analysed to identify usability issues that each participant encountered. For identifying usability issues from the screen recording data, user experience was evaluated by tracking resources used, events where the participant showed surprise, events where participant had to make difficult choices, context switches, misconceptions, difficulties faced, mistakes made, requested features and time taken for tasks. Cognitive Dimensions Framework [14] was used as a guidance in the analysis of both questionnaire responses and recordings. Usability issue reporting format introduced by Lavery et al. [34] was used to report issues. Finally, video recordings were analysed again to identify how usability issues that were identified, affected the participants for securely completing the task. Special attention was given to decisions made by participants that caused to reduce the security of programmes they developed.

# 4. Study Results

In this section, we are presenting results we obtained from this study. For the ease of presentation, we labeled participants with labels P1, P2,.., P10. They will be referred with this label from here onward. Statements made by participants that are presented in this section were not corrected for any grammatical errors and are presented as those were stated.

## 4.1. Security of the Developed Programmes

Table 2 shows the security of the programme each participant developed. Furthermore, it shows which components of the web application they successfully secured and which components they failed to secure.

We observed 3 main types of errors that participants made that resulted in them failing to secure the application. Those mistakes are,

1. Participants failed to identify all places in the source code that contained XSS vulnerability and hence fixed the code partially (P2, P7).
2. Participants used wrong encoding method to encode data. For example, used *encodeForHTML()* to encode data inside JavaScript (P2, P6).
3. Encoded input instead of output using one or two encoding methods. Mostly using *encodeForHTML()* (P3, P5, P8, P9, P10).

**Table 2. Security of Each Participant's Programme**

| Participant ID | Overall Security of the Aplication | Correctly Encoded HTML Content | Correctly Encoded HTML Attribute Content | Correctly Encoded Javascript Content |
|---|---|---|---|---|
| P1 | ✓ | ✓ | ✓ | ✓ |
| P2 | ✗ | ✓ | ✗ | ✗ |
| P3 | ✗ | Encoded input with encodeForHTML() | | |
| P4 | ✓ | ✓ | ✓ | ✓ |
| P5 | ✗ | Encoded input with encodeForHTML() | | |
| P6 | ✗ | ✓ | ✓ | ✗ |
| P7 | ✗ | ✓ | ✗ | ✓ |
| P8 | ✗ | Encoded input with encodeForHTML() | | |
| P9 | ✗ | Encoded input with encodeForHTML() | | |
| P10 | ✗ | Encoded input with encodeForHTML() | | |

## 4.2. Usability Issues of OWASP ESAPI

From the study, we could identify a total of 16 usability issues of OWASP ESAPI. Questionnaire responses revealed a total of 12 issues while video recording analysis revealed 12 issues. Each participant had encountered an average of approximately 7 usability issues. Here onward, we present each usability issue with the comments made by participants and what we observed.

Participants P2, P5, P6, P8 and P9 mentioned that there is too much information to read and learn in order to use the API to complete the task and fix XSS vulnerabilities. They mentioned that this makes it difficult for them to get the minimum understanding about the API that was required to use the API to fix XSS vulnerabilities. Furthermore, they mentioned that it made it difficult for them to understand which part of the API to use in order to achieve the goal. P6 mentioned in his response to the questionnaire that "*Even though the guidelines, the cheatsheet, was actually well written, the page itself was kinda long-ish.*". One of the main resource provided from OWASP about using OWASP ESAPI to fix XSS vulnerabilities is the "XSS Prevention Cheat sheet [8]" (At the time we conducted the experiment, this mainly referred to ESAPI. However, recently they have removed ESAPI specific code from this and made it a more generic one.[5]). Even though it provided all information required to use the API correctly to fix XSS vulnerabilities, most programmers did not follow it, saying it was too long. This made them refer to third party resources that provided partial solutions, which resulted in participants failing to successfully complete the task. This was elaborated by

[5] https://web.archive.org/web/20170615122701/https://www.owasp.org/index.php/XSS_(Cross_Site_Scripting)_Prevention_Cheat_Sheet

P3 as well where they mentioned that "*I did not even use the API documentation.There was a blog post with the required information, which was pretty much straight forward*".

Results from all participants showed that it is difficult to use the API without previous knowledge on some computer security related areas. Results revealed that it was hard to use the API without a previous knowledge on attacks, XSS attacks, XSS mitigation techniques and input sanitization. We could identify that lack of security knowledge specially made it hard to test whether the security vulnerability has been fixed properly, which resulted in participants falsely concluding that they have successfully fixed the XSS vulnerability. Some participants mentioned that they are not capable of testing what they developed due to their lack of security knowledge. P5 mentioned that "*It seems to me that output encoding untrusted data (which is done in one function call) sufficed, though I lack the knowledge to know if I've prevented all dangerous cases or not.*". Furthermore, P5 mentioned that API and its documentation did not provide details about XSS and it was difficult to find required information.

P1, P2, P3 and P8 mentioned that it was difficult to figure out how to achieve output encoding using the API. Some of them noted that programmer needs to have a previous knowledge on what type of encoding is required for each type of element. We observed that these participants spent a notable amount of time for searching things such as "*output encoding using OWASP ESAPI*" in Google and browsing through the results. P2 highlighted this in their response to the questionnaire saying "*It was bit hard to grasp the idea of what output encoding and how we can achieve this with the API.*".

P3, P5, P6, P7, P8 and P9 faced difficulties while completing the task due to issues in OWASP ESAPI documentation. Their results revealed that,

- Documentation is too lengthy.
- Documentation is not attractive for programmers.
- Documentation is difficult to understand.
- Some documentation is outdated.

These issues resulted in programmers referring to unreliable third party resources such as code samples in Stack Overflow to learn the API. This was observed in P7's recording and think-aloud results also where they mentioned that "*I tried to search for some documentation, but, I found out answers from stack overflow. Its really not bad*". Furthermore, above issues made it difficult for programmers to use the API and made the learning period lengthier. We observed that many participants spent a considerable time to learn the API by going through documentation and other resources. P7 explained their experience with the documentation saying "*I expected some documentation online - but after a few minutes I found that there is no up to date documentation available and I'd have to stick to JavaDoc and reading the sources*".

Another usability issue that participants encountered is the lack of examples. P2, P3, P6, P7 and P8 reported in their questionnaire responses as well as in their think aloud results that API lacks examples and a proper 'getting started' guide. P8 mentioned that "*In the document there are so many information but less usage examples. so it is hard to understand 1st time.*". Participants mentioned that lack of examples made it difficult for them to learn the API and also made it difficult to know what classes and methods of the API to use when writing code. Participants suggested that including examples into documentation will give a better experience to programmers. P6 suggested that "*I would have liked a big h1 sign saying "Example usage" for the APIs. Since I wasn't able to find examples as quick as I wished.*".

While completing the task, P5 and P10 found that Integrated Development Environment(IDE)'s suggestions are not working for the API. After entering method names, both these participants tried IDE suggestions to get an understanding about the required arguments for the method. Instead of providing useful suggestions, IDE showed parameter names such as arg0, arg1, etc. Participants mentioned that this made them spent more time for learning the API and it made it difficult to use the API.

P4 and P7 mentioned in their responses to the questionnaire that viscosity of the code that use the API was poor as it was difficult to make changes to the code that use the API. They mentioned that API had to be embedded into many places in the code and if they later find something that need to be changed, it need to be applied for many places, which made changing the code difficult. P7 elaborated saying "*There are too many places where ESAPI call is used - it'd be really hard if I'd be forced to alter the call ESAPI.encoder().whatever(value)*".

P1, P2, P3, P4, P5, P6, P9 and P10's results revealed that there were different methods that looked similar, but provided different functionalities and made it difficult for the participants to select correct methods to use in their code. These participants found it difficult to identify the difference between *encodeForHtml()* and *encodeForHtmlAttribute()* methods. This made it difficult for them to choose correct method to use, which sometimes resulted in them using a incorrect method and therefore, failed in fixing the vulnerability. P4 described issues they faced in their response to the questionnaire when we questioned about the consistency dimension [14] of the API. They said that "*There were kind of similar things. But if you are really familiar with html and other related technologies I think you can sort them after.For an example there are plenty of methods like encode content, encode attribute etc. Those may sound similar if you are not really familiar with html or any other related technologies*."

Results of P3, P4, P5, P8, P9 and P10 revealed that API did not provide any help to identify incorrect usages of the API. Results showed that API did not provide any help to identify that participants were using the API incorrectly, when they used it to encode input instead of output. Furthermore, API did not provide any way to inform participants when they used different methods in wrong contexts (eg: when using *encodeForHtml()* to encode values of Javascript contents). P3 elaborated on this in their questionnaire response when we asked if the API gave any help to identify that they used the API incorrectly. They mentioned "*no, I first used it to encode the input before saving, which was stupid*". However, P4 did not believe this as an issue of the API as they mentioned that *"Usually API itself doesn't give you clues of correct usages other than argument type mismatches.*".

Results from all participants revealed that end-user protection of the application would depend on the programmer who used the API to fix the vulnerability. Participants had to make choices while using the API that will affect the security of the code they developed and this could lead participants to develop less secure code using the API. P7 elaborated in their questionnaire response on how the security of the code would depend on the programmer. They mentioned that "*If I'd miss some place in JSP that prints out the user input, entire application would become vulnerable to XSS*". P5 suggested how API could minimize the dependency of security from programmers saying "*by providing a good level of abstraction and explanatory documentation, the API can help reduce the struggle of the programmer and minimize errors.*".

Results from P1, P3, P5, P6, P7 and P9 revealed that API did not provide any help for them to test the security of the code they developed using the API. This resulted in participants finding it difficult to verify whether they developed the code securely using the API or not. We observed that most of the participants did not properly tested the security of the code they developed even though the task was about fixing a security vulnerability. Most of them tested whether the vulnerability that could be exposed from HTML content has been fixed. But none of them tested vulnerabilities that could be exposed from HTML attribute and JavaScript content. It seemed that most of them did not have a sufficient knowledge to test whether the vulnerability has been fixed. While completing the task, P5 mentioned in their think-aloud output that "*I dont know if I finished the task. There might still be vulnerabilities in the application.*". Furthermore, P1 reported that it was difficult to evaluate the progress of a partially developed code that use the API, while completing the task. They elaborated saying "*I had to go through every page and find out what are the untrusted content in each page*".

## 5. Discussion

From the results of the study, we identified 3 types of mistakes that participants made that resulted in them failing to fix the vulnerability. We also identified 16 usability issues that were encountered by participants. We observed that some of these usability issues were a main reason for those mistakes that participants made.

To protect the application successfully, API required participants to identify all the places that the vulnerability existed and use the API to fix all these locations. However, P2 and P7 failed to identify all the places that the vulnerability was present and therefore resulted in only partially fixing the vulnerability. This made the API less effective and hence, less usable [35]. Some web frameworks such as Ruby 3.0 and React JS automatically escape inputs [1] without depending on the programmer to identify existing vulnerabilities and vulnerable code snippets. Integrating that sort of functionality, or a functionality to protect web applications from a more higher level (protect at web page level rather than going into element level) would enhance the usability of the API and hence, would help programmers to more effectively use the API to develop applications that are not vulnerable to XSS attacks.

Furthermore, we observed that P2 and P6 used wrong methods to encode data in some cases, which failed them in successfully completing the tasks. Some of the usability issues we identified aided participants in making this mistake. Lengthy documentation demotivated participants to properly read and grasp ideas in most cases. Therefore, most participants skimmed through documentation. When reading the "XSS Prevention Cheat Sheet" [8], most participants did not properly read it and hence, could not get a proper idea about different encoding methods required for different element types. However, P1 read "XSS Prevention Cheat Sheet" [8] completely before they started to fix the vulnerability. Their think-aloud results mentioned that they have previously used the API and hence, want to refresh their memory. It appeared in their think-aloud results that reading “XSS Prevention Cheat Sheet” [8] before starting to fix vulnerability was something they decided based on prior experience with the API. Other participants failed to identify that they need to read it and therefore, failed to get a proper idea about what encoding methods required for what elements. Furthermore, less attractiveness and lack of examples in official documentation made programmers referring to third party sources such as Stack Overflow. Those 3rd party sources mostly contained examples that use *EncodeForHTML()* method and did not highlight that different methods should be used to encode data, based on the type of element the data is used in. Therefore, participants used *EncodeForHTML()* method to encode data in HTML attribute and JavaScript elements that should have been encoded using different methods. Brief and attractive documentation with more examples would make it satisfactory and easy to follow those documentation and would help programmers to learn the correct way of using the API to protect their applications. It will also make the API less reliant on programmers’ previous knowledge on XSS attacks and XSS mitigation techniques, which would make the API easily usable for programmers, especially for those who are not experts of security.

Furthermore, participants blamed API as it did not provide any indication when they incorrectly used the API. Participants expected the API to notify them when they use an incorrect method to save a data element. They also expected that API would notify them or prevent them when they used the API to encode input data instead of output data that it is really supposed to encode. Previous research has also suggested that security APIs should be hard to misuse and they should notify programmers on incorrect use [14, 18]. It would be ideal for the API to provide better mechanisms to prevent programmers from incorrectly using it.

One of the main reasons for participants not been able to successfully fix the XSS vulnerability is that they did not properly tested whether the vulnerability has been fixed after they completed the task. It was apparent when observing think-aloud results and screen recordings of participants that most participants were not competent enough to test the security after applying the API and API did not helped them in this either. Some participants only tested with the sample malicious input we gave with task guidelines. Once an application worked fine with that input, they assumed that they have successfully fixed the vulnerability. Previous research has suggested that security APIs should help programmers to test the security of the applications that use the API [14, 17]. Therefore, OWASP ESAPI should either provide test routines and scripts for programmers to test their applications, or the API documentation should properly guide programmers on how to validate that they have properly used the API to remove XSS vulnerabilities in their applications.

## 6. Limitations

Since the main objective of our study was to identify usability issues of OWASP ESAPI, we used a pool of 10 participants. It has been acknowledged that a user study would identify about 80% of usability issues of an user interface by employing 4-5 subjects (i.e. participant users) [36]. Virzi identified that additional subjects are less likely to reveal new information and the most severe usability problems are likely to be detected in first few subjects [36]. Furthermore, more recently, Hwang and Salvendy [37] introduced “the 10±2 rule” where they argued that 10±2 users will be sufficient to conduct a usability evaluation. They also stated that small participant pools are adequate when using think aloud approach for usability evaluations. Even though this results are based on user studies conducted for user interfaces, our results suggest that this holds for API usability studies as well. We could identify 16 usability issues of the ESAPI and we could explain the mistakes that participants made based on these issues. However, because of the small sample pool we used, we could not infer any statistically significant results such as correlation between demographic variables and outcomes. We are planning to extend this study and explore these aspects in a future study.

## 7. Conclusion

In this study, we conducted a remote behavioural usability study with 10 software developers to identify usability issues that exist in OWASP ESAPI. Participants were asked to complete a simple programming task which required them to fix XSS vulnerabilities of a Java servelet web application. They had to think aloud and record their screens while completing the task and once they finished the task, they had to answer the cognitive dimensions based questionnaire [14]. Through the data we collected, we were able to identify usability issues that exist in OWASP ESAPI and how they affected the participants’ success in fixing the XSS vulnerability.

From the results we identified 3 types of mistakes that programmers did, which are,

1. Failing to identify all the places in the source code that contained XSS vulnerability.

2. Using wrong encoding method to encode data.

3. Using input encoding instead of output encoding.

These mistakes resulted them in failing to fix the XSS vulnerability properly. Furthermore, we identified 16 usability issues that exist in OWASP ESAPI. We observed that some of the usability issues we identified aided programmers into make mistakes that failed them in their task. Based on the results we proposed some improvements to OWASP ESAPI, which would enhance its usability and hence, would help programmers to more effectively and efficiently use it to fix XSS vulnerabilities.

## A. Sample page from the application given to participants

```
<body style="background-color:azure;">
<h1 align="center">Welcome to Forum!</h1>
<table align="center" width="95%" border="2">

    <tr>
        <td width="25%"> <h3>Subject</h3></td>
        <td width="25%"> <h3>Author</h3></td>
        <td width="25%"> <h3>Content</h3></td>
        <td width="25%"> <h3>Delete</h3></td>
    </tr>

    <%for(int i=0;i<posts.length;i++){%>

    <tr>

    <td title="<%=posts[i].getSubject() +
    "_subject"%>"><%=posts[i].getSubject()%>
    </td>

    <td title="<%=posts[i].getSubject() +
     "_author"%>"><%=posts[i].getAuthor()%>
     </td>

    <td title="<%=posts[i].getSubject() +
     "_content"%>"><a href=<%="article.jsp?id="
      + posts[i].getId()%>>View Content</a> </td>

    <td title="<%=posts[i].getSubject() + "_author"%>"
     align="center"><button type="button"
     onclick="deletePost('<%=posts[i].getSubject()%>',
     '<%=posts[i].getId()%>')">Delete</button></td>

    </tr>

    <%}%>

    <%if (posts.length <= 0){%>
    <tr>
        <th colspan="4">No Posts Found</th>
    </tr>
    <%}%>
    </table>
<br>
<a href="post.jsp" class="button">Create New Post
</a>
</body>
```